\documentclass[aps,superscriptaddress,eqsecnum,nofootinbib,showpacs,preprintnumbers]{revtex4-2}
\usepackage{amsmath}
\usepackage{amssymb}
\usepackage{amsfonts,amsthm,bm}
\usepackage{color,xcolor}
\usepackage[greek,english]{babel}
\usepackage{titlesec}
\usepackage{wrapfig}
\usepackage{graphicx,epsfig}
\graphicspath{{figures/}{./}}
\usepackage{float}
\usepackage{subfigure}
\usepackage{tikz}
\newcommand{\m}{\mu}
\newcommand{\n}{\nu}

\newcommand{\g}{\gamma}

\newcommand{\be}{\begin{eqnarray}}
	\newcommand{\ee}{\end{eqnarray}}
\newcommand{\bea}{\begin{eqnarray}}
	
	\newcommand{\eea}{\end{eqnarray}}

\def\a{\alpha}
\def\b{\beta}
\def\g{\gamma}

\def\la{\lambda}

\def\k{\kappa}
\def\m{\mu}
\def\n{\nu}
\def\r{\rho}

\def\th{\theta}

\def\Ref{\ref}

\newcommand{\beq}{\begin{equation}}
\newcommand{\eeq}{\end{equation}}
\newcommand{\bseq}{\begin{subequations}}
	\newcommand{\eseq}{\end{subequations}}

\begin{document}

{\leftline{KCL-PH-TH/2023-{\bf 06}}

\title{On the Thermal Stability of Hairy Black Holes}

\author{Nikos Chatzifotis }
\email{chatzifotisn@gmail.com} \affiliation{Physics Department, School of Applied Mathematical and Physical Sciences,
	National Technical University of Athens, 15780 Zografou Campus,
	Athens, Greece.}

\author{Panagiotis Dorlis}
\email{panos\_dorlis@mail.ntua.gr}\affiliation{Physics Department, School of Applied Mathematical and Physical Sciences,
	National Technical University of Athens, 15780 Zografou Campus,
	Athens, Greece.}

\author{Nick E. Mavromatos}
\affiliation{Physics Department, School of Applied Mathematical and Physical Sciences,
	National Technical University of Athens, 15780 Zografou Campus,
	Athens, Greece.}
\affiliation{Physics Department, King's College London, Strand, London WC2R 2LS, UK.}

\author{Eleftherios Papantonopoulos}
\email{lpapa@central.ntua.gr} \affiliation{Physics Department, School of Applied Mathematical and Physical Sciences,
	National Technical University of Athens, 15780 Zografou Campus,
	Athens, Greece.}

\begin{abstract}
We discuss thermodynamical stability for hairy black hole spacetimes, viewed as defects in the thermodynamical parameter space, taking into account the
backreaction of a secondary hair onto the spacetime geometry, which is modified non trivially. We derive, in a model independent way, the conditions for the hairy black hole with the secondary hair to reach a stable thermal equilibrium with the heat bath. Specifically, if the scalar hair, induced by interactions of the matter fields with quadratic-curvature corrections, produces an inner horizon in the deformed geometry, a thermodynamically stable configuration will be reached with the black hole becoming extremal in its final stage. We also attempt to make some conjectures concerning the implications of this thermal stability for the existence of a minimum length in a quantum space time. \color{black}
\end{abstract}

\maketitle

\flushbottom

\tableofcontents

\section{Introduction}\label{sec:intro}

Black holes are celestial objects within General Relativity (GR) and have been extensively studied in the literature, both from a classical and a quantum physics point of view. The recent observations of gravitational waves (GWs) \cite{Abbott:2016blz}-\cite{TheLIGOScientific:2017qsa}, which were generated during the collision of two black holes or neutron stars,
is leading to a new understanding of extreme gravity conditions.  During the final stages of massive stars, astrophysical black holes can be formed while in the early Universe density inhomogeneities  may produce primordial black holes.
Most  importantly, black holes can be viewed as thermodynamic configurations that have temperature and entropy and emit radiation from the black hole horizon.
The quantum effects in a curved spacetime are responsible for the Hawking evaporation mechanism, which has been studied extensively (for reviews see \cite{Unruh:1976db,Crispino:2007eb}). The quantum gravitational effects are expected to be dominant during the last stages of
the evaporation, when the semi-classical approach breaks down.

The stability of black holes is a central issue in GR and it has been studied for a long time starting from the pioneering work by Regge and Wheeler {\cite{Regge:1957a} and, in most cases, it was found that the black holes are stable. The most well known method  to study the stability of black holes is to calculate the quasinormal modes (QNMs) and their quasinormal frequencies (QNFs) \cite{Zerilli:1971wd}-\cite{Nollert:1999ji}. The calculation of QNMs and QNFs may give information about the
stability of matter fields that perturb a region outside  of a black hole, which, however, they do not back react on the metric. These perturbations allow for a probing mechanism of the black hole configuration, because they depend on the black hole charges (mass,
electromagnetic charge and angular momentum) and the fundamental constants of the gravitational theory.

Modified gravity theories, which are resulting from the presence of high curvature terms and from scalar fields coupled to gravity, known as scalar-tensor theories have been intensely studied. An important question in all such theories concerns their structure, the behaviour of black hole solutions and their stability. If the scalar field coupled to gravity backreacts on the background metric, then hairy black hole solutions can be generated in scalar-tensor theories \cite{Fujii}.  One of the first hairy black hole was discussed in \cite{BBMB}, but it was found that it was unstable \cite{bronnikov}. To cure this problem a new scale was introduced through a cosmological constant and hairy black hole solutions were found \cite{Martinez:1996gn}-\cite{Babichev:2013cya}.

Another type of modified gravity theories occur when a scalar field is directly coupled to a curvature topological term. One such theory is constructed by the coupling of a scalar field with the Gauss-Bonnet term in 4-dimensions \cite{stringT}, which allows for the existence of hairy black holes due to the violation of the energy conditions, bypassing in this way the no-hair theorem \cite{Kanti:1995vq}-\cite{Tang:2020sjs}. A second important example is the  Chern-Simons (CS) gravitational theory, that contains a non trivial (pseudo)scalar field coupled to the topological Pontryagin density, which is a consistent Lanczos-Lovelock gravity theory (for a review on  general CS gravities, see \cite{Zanelli:2005sa}). In \cite{CSblackhole}, black hole solutions with well defined AdS asymptotics were extracted in CS gravities, while in d-dimensional solutions, the stability of CS black holes under scalar perturbations were investigated in \cite{Gonzalez:2010vv}. CS gravity slowly rotating Kerr-like black hole solutions have been found and studied in \cite{Chatzifotis:2022mob,Chatzifotis:2022ene}.

Another interesting issue in modified gravities concerns the thermodynamic behaviour of black holes. Recently in \cite{Wei:2022dzw} the thermodynamical stability of black holes has been conjectured as a defect in thermodynamical parameter space. Specifically, by employing a generalized off-shell free energy, the authors of \cite{Wei:2022dzw} constructed appropriate vector fields in the black hole thermodynamical parameter space, which vanish for the specific solutions of various gravitational models, thus leading to the interpretation of the black  hole as a defect in the thermodynamic parametric space.\footnote{There is  a useful analogy from condensed matter physics, in which a defect is considered as the absence of a specific localised field excitation, e.g. a hole is considered as the ``absence'' of an electron at a certain spacetime point. Such an absence may be represented by the vanishing of an appropriate field quantity at that point, which indicates the corresponding vacancy implied by the ``absence'' of an electron.}
By working out some concrete examples, they
classified stable (unstable) black holes as  corresponding to appropriately defined positive (negative)
winding numbers corresponding to the defects. The  sum of the winding numbers for all the black hole branches at an arbitrary given temperature is found to be a universal  number (topological charge) independent of the parameters of the pertinent black hole geometry, but depends only on the thermodynamic asymptotic behavior of the black hole temperature at small and large black hole limits (compared to the Planck size black holes). Different black hole systems are characterized by three classes via this topological number (positive, zero and negative), which thus can be used for a better understanding  of the black hole thermodynamics, and, as the authors of \cite{Wei:2022dzw} conjecture, it may even shed light on such fundamental issues as the nature of quantum gravity, to which black holes are expected to play a fundamental role ({\it e.g.} in certain spacetime foam situations, microscopic black holes play a crucial r\^ole~\cite{Wheeler:1955zz,Wheeler:1998vs}). Verification of the conjecture of  \cite{Wei:2022dzw} has also been given for the case of rotating black holes of both Kerr (in d+1 spacetime dimensions) and Kerr-Newman (in (3+1) spacetime dimensions) type in \cite{Wu:2022whe}, including AdS rotating black holes~\cite{Wu:2023sue}, and also higher-curvature black holes of Lovelock~\cite{Bai:2022klw} or higher-dimensional GB AdS type~\cite{Liu:2022aqt}.

In this work, we first elaborate and explain the conjecture itself in its generality. Then, we go beyond the above black hole examples and examine black holes with secondary hair, back reacting on the black hole spacetime, which is thus deformed. Making use of the results of the conjecture, we determine the conditions under which such hairy black hole systems can be thermally stabilized, i.e. they are characterized by positive winding number, and hence leave a stable remnant. However, it is well known that exact analytic solutions in modified gravities with higher curvature couplings are not always accessible. To this end, we are presenting a model independent form of the hairy black hole metric with respect to a dimensionless parameter that controls the strength of the back reaction on the black hole geometry. We discuss the conditions that have to be satisfied, such that a thermodynamically stable black hole branch appears, which is due to the strength of the higher curvature interaction.

The validity of the form of the metric we are conjecturing is supported by several examples in (3+1)-dimensional spacetimes: one concerns a Schwarzschild black hole with dilaton secondary hair, in dilaton ($\phi$) GB (3+1)-dimensional higher-curvature gravitational theory with a linear in $\phi$ -GB interaction~\cite{Sotiriou:2013qea}, which admits perturbatively known analytic solutions. This is in contrast to the string-inspired Schwarzschild black hole in the case of an exponential in $\phi$-GB interactions~\cite{Kanti:1995vq}, where the solutions have only been determined numerically. This stringy case is also discussed here from the point of view of thermodynamical stability.
The second example, deals with rotating black holes in a string-inspired (3+1)-dimensional CS gravity~\cite{kaloper1,kaloper2,kaloper3,jackiw,yunes} with axion secondary hair back reacting on the geometry, which thus deforms the slowly rotating Kerr solution~\cite{Chatzifotis:2022mob}. Such deformations are only formally known, through recursive relations order by order in perturbation theory.
A final example is the (3+1)-dimensional extended GB gravity \cite{Fernandes:2021dsb}, which includes coupling of a conformal scalar to a GB curvature combination.

All examples point non-trivially to the fact that the strength of the back reaction is controlled by a dimensionless parameter, which is the ratio of the dimensionful higher curvature coupling with respect to the black hole size. As such, when the black hole decreases in size and becomes comparable with the length scale that the coupling constant introduces, the back reaction terms cease being subdominant and may in principle stabilize the thermodynamic fate of the black hole.

The work is organized  as follows: in section \ref{sec:bath} we review the behaviour of a black hole embedded in an external thermal bath, corresponding to a temperature $T$, and present thermodynamical stability arguments, which will hopefully shed more light to the stability criterion of \cite{Wei:2022dzw}. In section \ref{sec:topol} we review the work of \cite{Wei:2022dzw} and verify the conjecture that associates the local thermodynamical stability with an appropriate topological charge. In section \ref{sec:inter}, we focus on hairy black holes, exploring the conditions under which the interaction between Gravity and matter fields may lead to a stable black hole branch of positive winding number.  We also make conjectures on the implications of such stability on the existence of a potential minimal length in the quantum gravity space time.
Finally, section \ref{sec:concl} contains our conclusions and outlook.

\section{ Black Holes in a Thermal Bath and Stability Arguments}
\label{sec:bath}

In this Section we will review the thermodynamic stability of asymptotically flat black holes. One can distinguish two thermodynamic regimes that affect the stability properties of a black hole, henceforth denoted as high and low temperature regimes. It is known from statistical physics, that the high temperature regime can be treated classically, in the sense that such systems have vanishingly small quantum contributions. In the high temperature regime, we can have a fully classical framework in which the black hole is large and a semi-classical regime for small black holes.

To better understand the role of the temperature to the stability of a black hole, we consider a black hole in a heat bath which means that the total system  is in thermal equilibrium with temperature $T$. This leads to the following thermodynamic statement
\begin{equation}
	T_{BH}=T~,
	\label{temptStatPhys}
\end{equation}
indicating that  the black hole can be described in terms of the canonical ensemble.  On the other hand, we know from the black hole physics that the temperature of a black hole corresponds to a geometric quantity,  the surface gravity, $\k_g$,
\begin{equation}
	T_{BH}=\frac{\k_g}{2\pi}~.
	\label{tempBhPhys}
\end{equation}
Combining equation \eqref{temptStatPhys} and \eqref{tempBhPhys}, we get
\begin{equation}
	\k_g=2\pi T~,
	\label{BhStatPhys}
\end{equation}
which combines a geometric quantity of the black hole (surface gravity) with the temperature of the heat bath $T$, which can vary freely and independently of the black hole.

As we consider the canonical ensemble, charges of the black hole beyond the mass are assumed to be fixed. Thus, we can express the surface gravity as a function of $r_h$, $\k_g=\k_g(r_h)$. This means that $r_h$ plays the role of an independent parameter that characterizes the black hole by its size. In this sense, equation \eqref{BhStatPhys} corresponds to an equation that relates the size of the black hole and the temperature of the heat bath. For different black holes, the surface gravity has different dependence on $r_h$, which consequently leads to different configurations for black holes that can be in thermal equilibrium with the heat bath. For example, we know that the surface gravity of the Schwarzschild black hole is given by
  \begin{equation}
	\k^{Schw}_g=\frac{1}{2r_h}~.
\end{equation}
In view of \eqref{BhStatPhys} the event horizon of the Schwarzschild black hole is related with temperature of the heat bath, by the following equation
\begin{equation}
	\tau=4\pi r^{Schw}_h~,
	\label{SchwBranch}
\end{equation}
which is a linear dependence between $\tau$ and $r_h$, where $\tau=T^{-1}$,  with dimensions of time. Thus, assuming the Schwarzschild black hole in a given temperature via an appropriate heat bath, there exists only one size for the event horizon, i.e. there is only one size configuration available for the black hole.
 On the other hand, the surface gravity of the Kerr-Newman black hole is given by
 \begin{equation}
	\k^{KN}_g=\frac{r_{KN,h}-r_-}{2(r_{KN,h}^2+\a^2)}~, \quad r_-=M-\sqrt{M^2-Q^2-\a^2}\,,
\end{equation}
where $\a$ is the rotation parameter (angular momentum per unit mass). Combining again with \eqref{BhStatPhys}, we have
\begin{equation}
	\tau=4\pi\frac{r_{KN,h}^3+\a^2 r_{KN,h}}{r_{KN,h}^2-Q^2-\a^2}~,
\end{equation}
which is a cubic equation, meaning that, in general, there is not a 1-1 correspondence between the size of the black hole and a given heat bath temperature. For the Reissner-Nordstrom $(\a=0)$ and the Kerr black hole $(Q=0)$ the above reduces to the following expressions
\begin{align}
	\label{RNBranches}
	&\tau=\frac{4\pi r_{RN,h}^3}{r_{RN,h}^2-Q^2}:\;\;\text{Reissner-Nordstrom}\\
	&\tau=4\pi\frac{r_{K,h}^3+\a^2 r_{K,h}}{r_{K,h}^2-\a^2}:\;\;\text{Kerr}
	\label{KerrBranches}
\end{align}

To understand the physical significance of the relation of the temperature and the size of the black hole, we plot in Fig.~\ref{bhbranches} the relations between $\tau$ and $r_h$ for different black holes corresponding to the equations \eqref{SchwBranch}, \eqref{RNBranches}, \eqref{KerrBranches}, from which we can distinguish the possible branches of black holes for each specific case. As we already mentioned, the Schwarzschild black hole in a given heat bath temperature yields only one possible size. Beyond this, we see that the introduction of charge (Reissner-Nordstrom) or rotation (Kerr) introduces two branches, one for small and one for larger black hole configurations.
\begin{figure}[h]
\centering
\includegraphics[width=.35\textwidth]{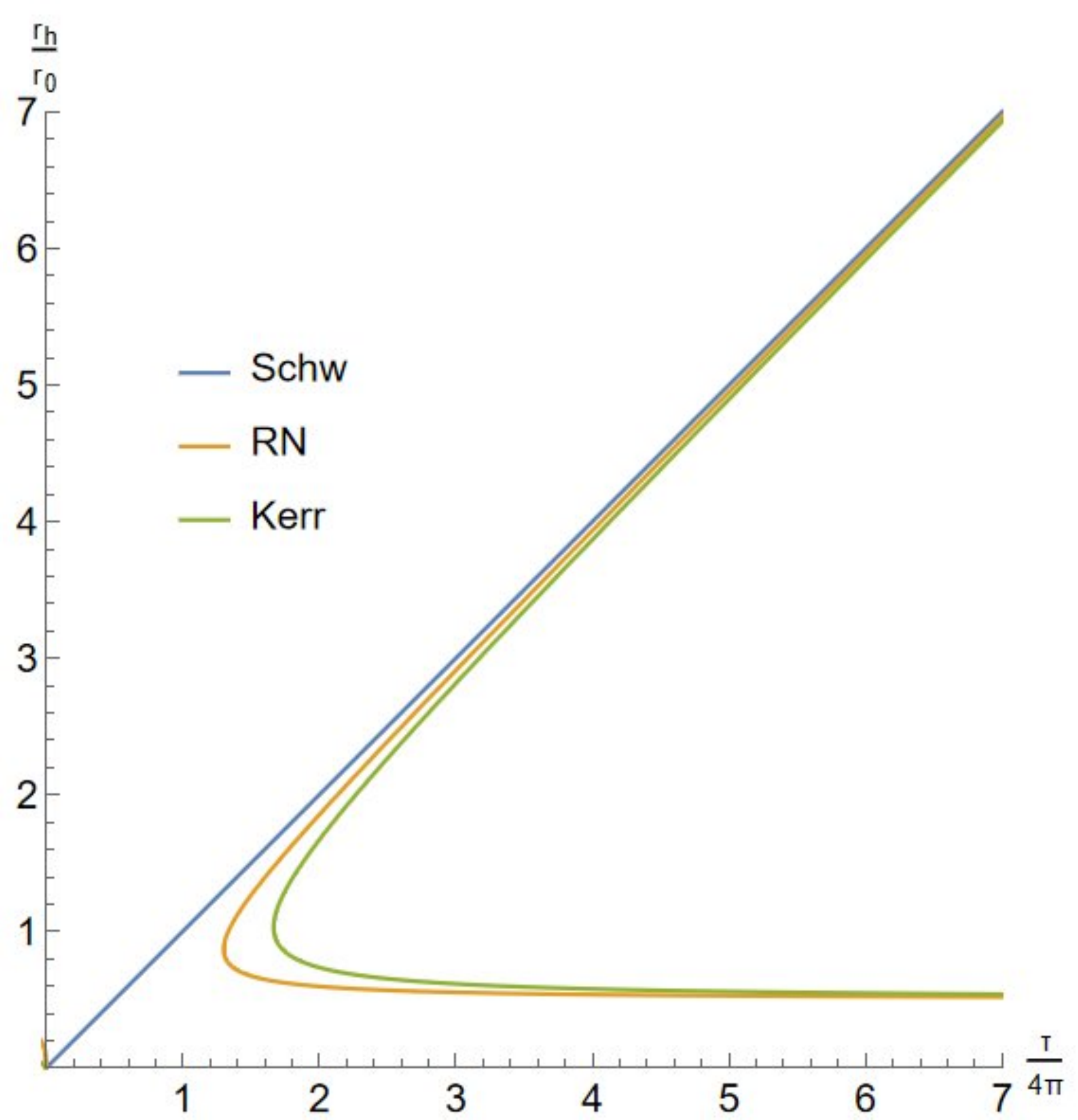}
\caption{Branches of static, stationary and charged black holes in General Relativity.}
 \label{bhbranches}
\end{figure}
According to these branches, there is an upper bound for the temperature, above which thermal equilibrium is impossible to be achieved. Moreover, there is also a lower bound for the size of the black hole, beyond which a black hole cannot exist in thermal equilibrium. Of course, the corresponding critical values for the temperature and the size of the black hole depend on the details of the black hole system that is considered. However, it is of great interest to investigate under which conditions the above qualitative behaviour can appear.

To  investigate the stability of the black hole we observe from Fig.~\ref{bhbranches}  that the large black hole branch corresponds to a negative slope of $r_h$,  $\partial r_h/\partial T<0$, while the small one to a positive one, $\partial r_h/\partial T>0$. The crucial quantity that help us to investigate the local stability issue is the heat capacity which is defined by
\begin{equation}
	\mathcal{C}_Q=T\frac{dS}{dT}~,
\end{equation}
and by the chain rule can be written as
\begin{equation}\label{cq2}
	\mathcal{C}_Q=T\frac{\partial r_h}{\partial T}\frac{\partial S}{\partial r_h}~.
\end{equation}
Since, the entropy increases with increasing surface, its derivative w.r.t  $r_h$ has to be positive-definite, which consequently means that the sign of the heat capacity is determined by the sign of  $\partial r_h/\partial T$. Thus, in general, for the system of two branches of the above form:
\begin{itemize}
	\item Large black hole branch: $\mathcal{C}_{Q}<0\rightarrow$ Unstable branch
	\item Small black hole branch: $\mathcal{C}_{Q}>0\rightarrow$ Stable branch
\end{itemize}
since $\partial r_h/\partial T<0$ for the large black hole branch and $\partial r_h/\partial T>0$ for the small one.

Let us consider a large system of temperature $T$ with all of its parts in thermal equilibrium. Deviations $\pm\delta T$ from the temperature of the heat bath $T$ might appear in small parts of the system, in the form of thermal fluctuations as can be seen in Fig.~\ref{thermal}. In such a case, a heat transfer occurs between the total system and the irregularities from the hot to the cold. A system with positive heat capacity decreases/increases its temperature, when it emits/absorbs thermal energy, while a system with negative heat capacity behaves conversely. Thus, when small temperature deviations appear in parts of positive heat capacity, the deviation will fastly vanish and equilibrium is recovered. On the other hand, when small temperature deviations appear in parts of negative heat capacity, the deviation grows up and the thermal equilibrium breaks down. In this sense, the parts of positive/negative heat capacity are in a stable/unstable thermal equilibrium with the heat bath, depending on the evolution of small perturbations in temperature.
\begin{figure}[h]
\centering
\includegraphics[width=.6\textwidth]{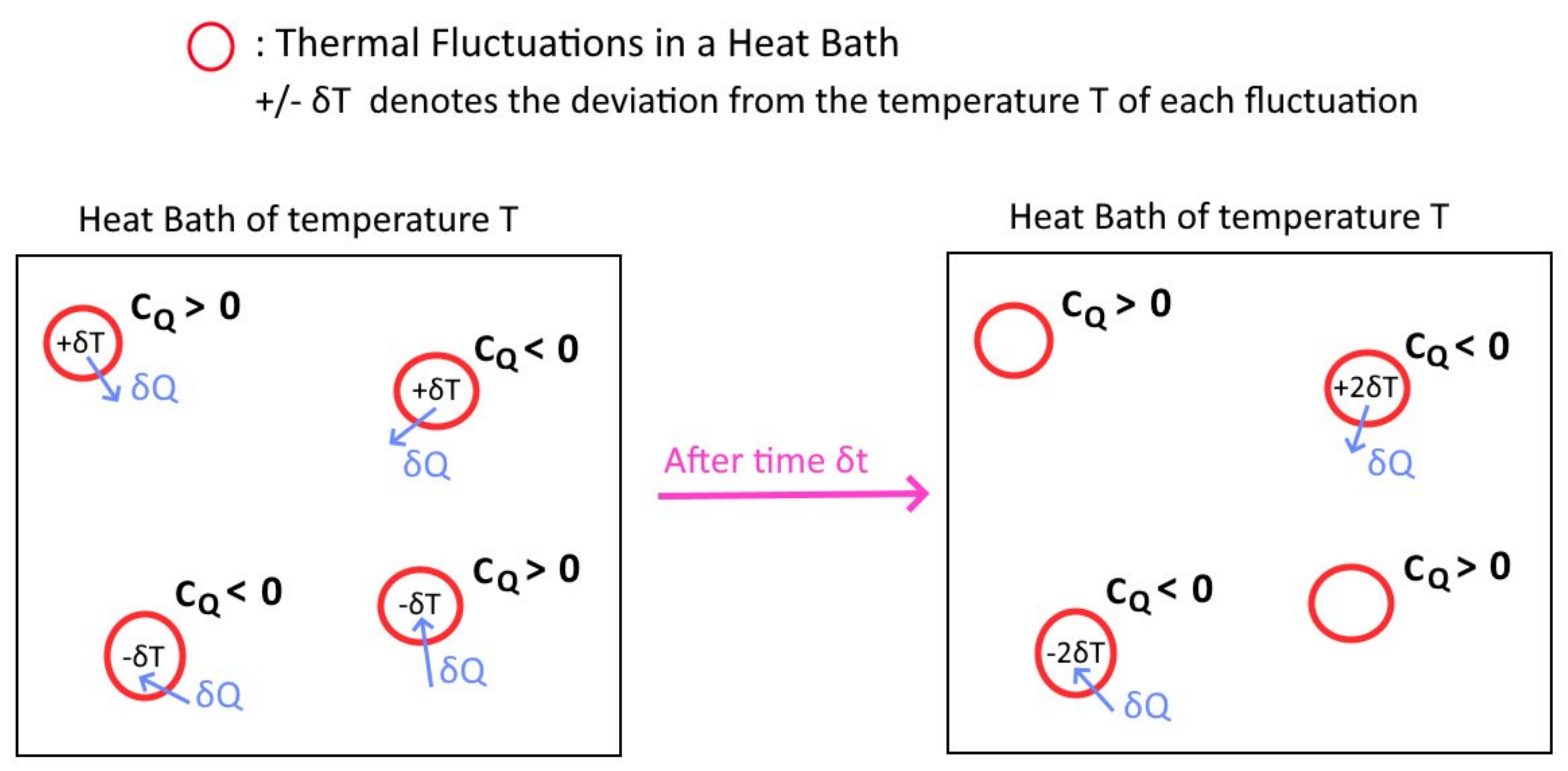}
\caption{Initially, all the parts of the system are in thermal equilibrium, with the same temperature $T$. When thermal fluctuations appear, deviations in temperature will appear in small parts of the system. Small temperature deviations for parts of positive heat capacity will vanish fast implying thermal stability. On the other hand, small temperature deviations for parts of negative heat capacity are enhanced, thus leading to thermally unstable configurations.  }
 \label{thermal}
\end{figure}

Stable and unstable thermal equilibrium is also related to the free energy of the canonical ensemble, which is defined as follows
\begin{equation}
	\mathcal{F}=E-\frac{\mathcal{S}}{\tau}~,
	\label{FreeEnergy}
\end{equation}
where $E$ is the total energy of the system, $\mathcal{S}$ is the entropy\footnote{For a black hole, the total energy of the system corresponds to the ADM mass, while the entropy in an arbitrary diffeomorphism invariant theory is given by ther well known Wald's formula \cite{Wald:1993nt}.} and $\tau$ is the freely varying parameter introduced previously, $\tau=T^{-1} $. Since we consider the canonical ensemble, every charge beyond the mass $M$ of the black hole is considered fixed. Thus, we can consider $E$ and $S$ and consequently $\mathcal{F}$ as functions only of $M$, with electric charge and angular momentum or any other charge as fixed. This is equivalent as to consider the above quantities as functions of the black hole radius $r_h$, as we already stated.

By construction, the free energy is defined in such a way that for any transition of the system, from one state to another, its free energy cannot increase \cite{fermi1956thermodynamics}. This means that the minima of the free energy correspond to a stable state, while its maxima to an unstable one. The free energy is expressed in terms of $r_h$ and thus its extremum points are determined by its derivatives, $r_h$. Thus, at an $r_h=R$, where $\partial\mathcal{F}/\partial r_h\vert_{r_h=R}=0$, we have
\begin{enumerate}
	\centering
	\item $\displaystyle \frac{\partial^2\mathcal{F}}{\partial r_h^2}\Big\vert_{r_h=R}>0\rightarrow$  Stable equilibrium\\
	\item $\displaystyle \frac{\partial^2\mathcal{F}}{\partial r_h^2}\Big\vert_{r_h=R}<0\rightarrow$  Unstable equilibrium\\
\end{enumerate}
The above stability statements in terms of the free energy are related to perturbations of $\mathcal{F}$,  $r_h$. However, in view of Fig.~\ref{thermal}, perturbations of $r_h$ can be realized as perturbations of the temperature. In this sense, the stability criteria that we previously present in terms of the heat capacity and the temperature correspond to those we stated  to the free energy and the size of the black hole.  This can be easily verified from Fig.~\ref{fig:FreeEnergy}, in which we see the free energy of the Schwarzschild, Reissner-Nordstrom and Kerr black holes for a given temperature $T$, for which the two branches appear. As we can see the branch of the small black hole corresponds to a local minimum, while the large black hole to a local maximum, in agreement with the previous discussion.
\begin{figure}[h!]
	\centering
	\includegraphics[width=0.5\textwidth]{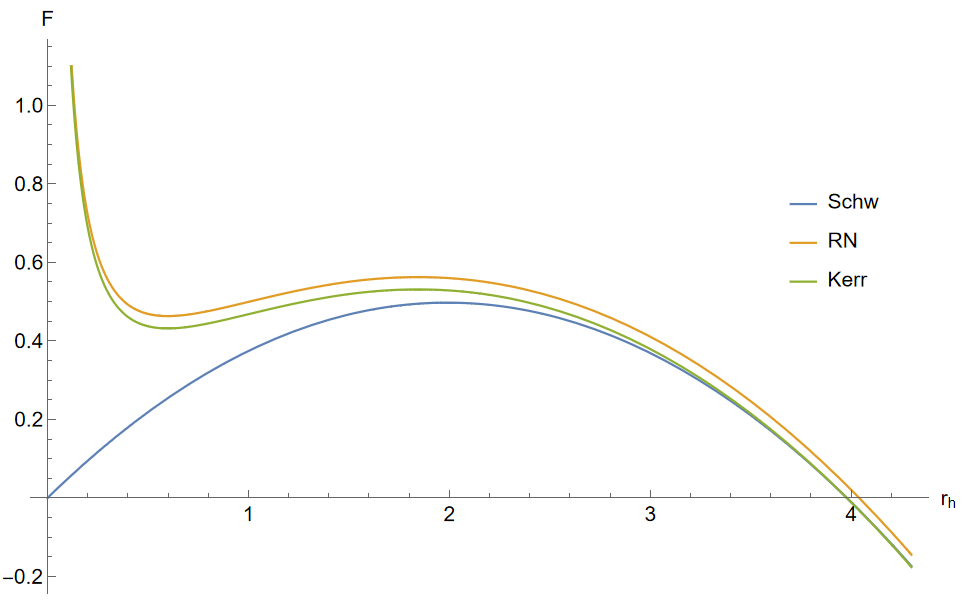}
	\caption{The Free Energy for static, stationary and charged black holes in General Relativity.}
	\label{fig:FreeEnergy}
\end{figure}

\section{Black Holes as Thermodynamical Topological Defects }
\label{sec:topol}

  We are now well equipped to review in this section the topological approach 
of \cite{Wei:2022dzw} for investigating the different branches of black holes in thermal equilibrium with the heat bath. Specifically, the properties of the black hole solutions are studied through the notion of topological defects, that is, zero points of some vector field, $\Phi^a(\vec{x})$. The defects are being thermodynamic because the field $\Phi^a(\vec{x})$ is defined through thermodynamic parameters of the black hole. Precisely, the generalized free energy of the canonical ensemble is introduced as in \eqref{FreeEnergy}. Then, the field $\Phi^a$ is constructed as follows
\begin{equation}
	\Phi^a=\left(\Phi^{r_h},\Phi^\Theta\right)=\left(\frac{\partial\mathcal{F}}{\partial r_h},-\cot\Theta\csc\Theta\right)~,
\end{equation}
where $r_h$ is the location of the event horizon and $\Theta$ is an auxilliary parameter defined in $0\leq\Theta\leq\pi$. With this choice for $\Phi^\Theta$, the zeros of $\Phi^a$ always lie on the horizontal axis $\Theta=\pi/2$ of the $r_h-\Theta$ plane. Thus, the defects are specified by $\Theta=\pi/2$ and $\partial\mathcal{F}/\partial r_h=0$. The parameters of the black hole that are introduced are its horizon, $r_h$, and the black hole temperature $T_{BH}$. These defects correspond to the black hole, since the condition $\tau=T^{-1}_{BH}$ is  equivalent with the condition $\partial\mathcal{F}/\partial r_h=0$. For $\tau\neq T^{-1}_{BH}$ the generalized free energy is off-shell, i.e. it does not correspond to a solution of the Einstein equations. Thus, by taking $\tau=T^{-1}_{BH}$, we set the free energy on-shell and consequently $\Phi^a=0$ correspond to the zeros of the tensor field $\mathcal{E}_{\m\n}=G_{\m\n}-\k^2T_{\m\n}=0$. In this sense, the roots of $\Phi^a$ correspond to the black hole solutions of $\mathcal{E}_{\m\n}$. In order to proceed, the following topological current is introduced~\cite{Duan:1979ucg,duan2}
\begin{equation}\label{tc}
	j^\m=\frac{1}{2\pi}\epsilon^{\m\n\r}\epsilon_{ab}\partial_{\n}n^a\partial_\r n^b\;\;\text{where}\;\;\m,\n,\r=0,1,2\;\;\text{and}\;\;a,b=1,2~,
\end{equation}
with the notation $\partial_\n=\partial/\partial x^\n$ for $x^\n=(\tau,r_h,\Theta)$, while $\epsilon$ corespond to the Levi-Civita symbols in the appropriate spaces. Moreover, the unit vector $n$ is defined as $n^a=\Phi^a/\vert\vert\Phi\vert\vert$, where $\Phi^{1,2}=\Phi^{r_h,\Theta}$. The presence of the Levi-Civita symbol in the topological current makes it by definition divergenceless (conserved), $\partial_\m j^\m=0$, which allows us to define a \emph{topological charge} as
\begin{equation}
	Q=\int_\Sigma d^2x j^0~,
	\label{TopologicalCharge}
\end{equation}
where $\Sigma$ is some region on the parameter space $(r_h,\Theta)$, while its boundary, $\partial\Sigma$, is a curve in the $r_h-\Theta$ plane. Following~\cite{duan2}, one can show that the topological current \eqref{tc} reads
\begin{equation}
	j^\m=  \delta(\vec{\Phi}) \, J^\mu \Big(\frac{\Phi}{x}\Big), \quad J^\mu \Big(\frac{\Phi}{x} \Big) \equiv  \frac{1}{2} \,\epsilon_{ab} \, \epsilon^{\mu\nu\rho} \partial_\nu \Phi^a \, \partial_\rho \Phi^b \,,
\label{TopologicalCurrentJacobi}
\end{equation}
where we used that the two-dimensional Laplacian $\Delta_{\Phi^a} \, {\rm ln}||\vec \Phi|| = 2\pi \, \delta (\vec \Phi)$. Eq. \eqref{TopologicalCurrentJacobi} means that the topological current is non-zero only at the zeros of $\Phi^a$. Thus, the topological charge in non-zero only for regions, $\Sigma$, in the parameter space that include a defect. If $\Sigma$ does not include any defect, then $Q=0$. Thus, as we have attached every defect with a black hole solution, this topological charge is attached with the corresponding black holes. Moreover, it turns out that\footnote{For more details about the proof of these statements, see \cite{Wei:2022dzw} and \cite{Wei:2020rbh}.}:
\begin{equation}
	Q=\sum_{i=1}^N w_i~,
\end{equation}
where $i=1,2,...,N$ counts for the defects that $\partial\Sigma$ encloses and $w_i$ is the winding number of $\partial\Sigma$ for the i-th defect. Consequently, we can define a local and a global topological charge depending on the curve in $r_h-\Theta$ that we choose to be the boundary, $\partial\Sigma$, of $\Sigma$. Suppose that there are $N$ zeros of $\Phi^a$. If $\partial\Sigma$ encloses only the i-th zero of $\Phi^a$, then the topological charge accounts only for the i-th defect and specifically is equal to $w_i$. This is the \emph{local topological charge}. If $\partial\Sigma$ encloses all of the parameter space, then accounts for all the zeros of $\Phi^a$. This is the \emph{global topological charge}.
However, if someone calculates the winding number for each defect, then, in order to find the global topological number, has just to sum all the winding numbers of each defect.

The correspondence between the topological charges and the thermodynamic stability of each branch is conjectured in \cite{Wei:2022dzw}. Namely, it is conjectured that branches with winding number $+1 \, (-1)$ correspond to a stable (unstable) branch, respectively.\footnote{Of course, the winding number depends on the orientation of the boundary $\partial\Sigma$. So, we follow the convention introduced in \cite{Wei:2022dzw} and we always refer to counter-clockwise curves.}  As we have mentioned previously, local stability of each branch is determined by the sign of $\partial^2\mathcal{F}/\partial{r_h}^2$. So in \cite{Wei:2022dzw}, it is conjectured that the winding number of each defect is related to the sign of $\partial^2\mathcal{F}/\partial{r_h}^2$.  In other words, the afforementioned conjecture states that
\begin{equation}
	w_i={\rm sgn}\left(\frac{\partial^2\mathcal{F}}{\partial r_h^2}\Bigg\vert_i\right)~.
	\label{conjecture}
\end{equation}
In order to verify this, one has to notice that the information of $\partial^2\mathcal{F}/\partial{r_h}^2$ lies on the derivative $\partial \Phi_{r_h}/\partial r_h$. Then, if we substitute \eqref{TopologicalCurrentJacobi} into \eqref{TopologicalCharge}, we get
\begin{equation}
	\begin{aligned}
		&Q=\int_\Sigma dr_h\;d\Theta\;\delta(\Phi_{r_h})\;\delta(\Theta)\;J^0\left(\frac{\Phi}{x}\right)
	\,\,	\Rightarrow \,\, &Q=\int_\Sigma dr_h\;d\Theta\;\delta(\Phi_{r_h})\;\delta(\Theta)\;\frac{\partial\Phi_{r_h}}{\partial r_h}\;\frac{\partial\Phi_\Theta}{\partial\Theta}~.
	\end{aligned}
\end{equation}
Recall the identity of the delta function
\begin{equation}
	\delta(f(x))=\sum_i\frac{\delta(x-z_i)}{\vert\frac{df}{dx}\vert_{z_i}}~,
\end{equation}
where $z_i$ are those that $f(z_i)=0$. Then, we can write
\begin{equation}
	\label{topcharge}
	\begin{aligned}
		Q=\left(\Bigg\vert\frac{\partial\Phi_\Theta}{\partial\Theta}\Bigg\vert_{\Theta=\pi/2}\right)^{-1}\int d\Theta& \delta(\Theta-\pi/2)\frac{\partial\Phi_\Theta}{\partial\Theta}\sum_i\left(\Bigg\vert\frac{\partial\Phi_{r_h}}{\partial r_h}\Bigg\vert_{r_h=z_i}\right)^{-1}\int dr_h \delta(r_h-z_i)\frac{\partial\Phi_{r_h}}{\partial r_h}\\
		\Rightarrow\; \, &Q= \sum_i {\rm sgn} \left( \frac{\partial \Phi_{r_h}}{\partial r_h}  \Bigg\vert_i \right)~,
	\end{aligned}
\end{equation}
since the only zero of $\Phi_\Theta$ lies at $\Theta=\pi/2$ and $\partial\Phi_\Theta/\partial \Theta\vert_{\pi/2}=1>0$. Moreover, if $\Sigma$ includes only the i-th defect, then the above sum reduces to the local topological charge of the corresponding defect, $w_i$. In addition, since $\Phi_{r_h}=\partial\mathcal{F}/\partial r_h$, we have that $\partial\Phi_{r_h}/\partial r_h=\partial^2\mathcal{F}/\partial r_h^2$, which implies \eqref{conjecture}, and therefore the conjecture appears to be true. Having verified the conjecture, we shall now proceed with the analysis of hairy black hole spacetimes and test whether the secondary hair components may lead to a stable black hole branch, which, as follows from the previous discussion, is assigned to a positive winding number \eqref{conjecture}.
\\
\\
\\

\section{Back Reaction of Secondary Hair and Black Hole Thermodynamical Stability}\label{sec:inter}

In this Section, we will investigate the conditions, under which black hole thermal stability can be achieved as a result of an effective interaction between the matter and the gravitational field, which in turn dresses the black hole with secondary hair. To this end, we first note that
the Einstein-Hilbert action can be considered as the action of the free gravitational field. Considering only spherically symmetric black holes, the Schwarzschild solution describes the black hole of the free gravitational theory. We shall be interested in interactions of the gravitational field with matter fields that have the  form of a higher curvature coupling, with the strength of the interaction measured by a dimensionfull coupling constant $A$. We can write the general Lagrangian as
\begin{equation}
	\mathcal{L}=\mathcal{L}_{EH}+\mathcal{L}_{matter}+A \, \mathcal{L}_{int}~,
	\label{InteractingLagrangian}
\end{equation}
where $\mathcal{L}_{EH}$ is the Einstein-Hilbert action, $\mathcal{L}_{matter}$ the kinetic terms for the matter fields, while $\mathcal{L}_{int}$ accounts for the possible interactions between the matter fields and the gravitational field. Working in natural units where $\hbar=c=1$, all the physical quantities are reduced into units of length or energy. In this respect, the introduction of a dimensionfull coupling constant introduces a new length scale that characterizes the interaction of matter with gravity; if it is not itself of length dimension, the length scale can be defined by some power of it in order to acquire the appropriate dimension. We will keep the coupling constant, $A$, to be of length dimension for notational convenience.

It is well known that if such interactions are present, they may produce non-trivial hairy black hole spacetimes, thus violating the no-hair theorem. However,  the coupling constant $A$ which  denotes the strength of the interaction is not the only parameter that determines the impact that such an interaction has in a black hole. This is because a black hole has a characteristic scale determined by its event horizon radius, $r_h$. We claim that the impact of an interaction to a black hole can be realized by the dimensionless parameter $\gamma$
\begin{equation}\label{coupling}
	\gamma= \frac{A\k}{r_h^2}~.
\end{equation}
This parameter contains the coupling constant of the interaction $A$, the Planck length (through $\k$) and the size of the black hole, through $r_h$ . For a fixed coupling constant, $\gamma$ becomes larger for smaller black holes. This means, that as the black hole shrinks (for example, through Hawking radiation), its interaction with the matter fields becomes more and more important and especially with an inverse square relation of its radius. Thus, in the large black hole regime the interactions can be treated perturbatively, in the sense that we are able to keep terms up to first order in $A$ and consequently in $\g$. In this sense, one finds perturbative solutions for the black hole around the solution of the free gravitational field. However, this is not true for black holes of small size because the effects of the interaction are no longer subdominant and a perturbative expansion to first orders is not valid.

\subsection{ Stable Black Holes with Secondary Hair}

Let us give some examples in order to proceed into some general considerations about the backreaction that such interactions have on a black hole metric. Firstly, consider a scalar field coupled linearly with the GB topological term given by the following action
\begin{equation}
S=\int d^4x\sqrt{-g}\left[ \frac{R}{2\k^2}-\frac{1}{2}(\partial\phi)^2 +A \phi\, \mathcal{R}_{GB} \right]~,
\label{GaussBonnetAction}
\end{equation}
where $\mathcal{R}_{GB} \equiv R_{\mu\nu\rho\sigma}\, R^{\mu\nu\rho\sigma} - 4\, R_{\mu\nu}\, R^{\mu\nu} + R^2 $ denotes the GB term, which is a topological  term in (3+1) dimensions. The coupling constant has itself the right dimensions, i.e. it has dimensions of length. It is known that hairy spherically symmetric black holes exist in the non-linear exponential dilaton coupling case \cite{Kanti:1995vq}, while an analytic solution for the linear coupling (weak dilaton approximation) has been found only perturbatively \cite{Sotiriou:2013qea}.\footnote{In the (3+1)-dimensional string-inspired case of \cite{Kanti:1995vq}, the GB action with a canonically normalised dilaton field of mass dimension $+1$, is given by
\begin{align}\label{strings}
\mathcal  L_{int} = \frac{\alpha^\prime}{8\, \kappa^2\, g^2_s} \, e^{-\kappa \phi/\sqrt{2}} \, \mathcal{R}_{GB} \,,
\end{align}
with $g_s$ the string coupling, and $\alpha^\prime=1/M_s^2$ the Regge slope, with $M_s$ the string mass scale. The zeroth-order term in a weak dilaton expansion is a total derivative and does not contribute to the action. If we restrict ourselves to classical dilaton fields that assume subplanckian values $\phi \kappa \ll 1$, then one recovers the action \eqref{GaussBonnetAction} by truncating the expansion in powers of $\phi$ to first order.}
The solution for the $g_{tt}$ component, up to second order in $A$ (equivalently in $\gamma$) takes the following form
\begin{equation}
g_{tt}=-1+\frac{r_h}{r}+\g^2\left[\frac{20}{3}\left(\frac{r_h}{r}\right)^7-\frac{16}{5}\left(\frac{r_h}{r}\right)^6-\frac{22}{5}\left(\frac{r_h}{r}\right)^5-\frac{52}{3}\left(\frac{r_h}{r}\right)^4-\frac{4}{3}\left(\frac{r_h}{r}\right)^3+\frac{49}{5}\frac{r_h}{r}\right] +\mathcal{O}(\g^4) ~,\\
\label{gttorder2gaussbonnet}
\end{equation}
where $r_h$ is the radius of the event horizon. \par
As a second example, we present the dynamical Chern-Simons gravitational theory, obtained by the following action
\begin{equation}
S=\int d^4x\sqrt{-g}\left[ \frac{R}{2\k^2}-\frac{1}{2}(\partial\phi)^2 -A\phi\mathcal{R}_{CS} \right]
\label{ChernSimonsAction}
\end{equation}
where $\mathcal{R}_{CS} = R^{\mu\nu\rho\sigma} \, \widetilde R_{\mu\nu\rho\sigma}$, with
$ \widetilde R_{\mu\nu\rho\sigma} = \frac{1}{2} R^{\,\,\,\,\,\,\,\alpha\beta}_{\mu\nu}\varepsilon_{\alpha\beta\rho\sigma}\, $
the dual Riemann tensor, and $\varepsilon_{\mu\nu\alpha\beta}$ the covariant Levi-Civita symbol,
 is the Chern-Simons term, which in (3+1) dimensions is also topological.\footnote{In
generic Chern-Simons gravity~\cite{jackiw,yunes}, $A=1/f_\phi$, where $f_\phi$ is the axion coupling
with mass dimensions $+1$, while in string-inspired Chern-Simons gravity~\cite{kaloper1,kaloper2,kaloper3} of the so-called gravitational axion, which in (3+1) dimensions is dual to the field strength of the spin-1 antisymmetric  tensor field of the massless gravitational string multiplet, we have
\begin{align}\label{stringsCS}
A = \sqrt{\frac{2}{3}}\frac{\alpha^\prime}{48\,\kappa}\,.
\end{align}} This interaction term is turned on only for rotating black holes, because $\mathcal{R}_{CS}$ vanishes in the presence of spherical symmetry. However, the features of the solution that we are interested at this point are independent of this fact; we are interested only on the form of the backreaction. Analytical solution for this interaction is possible only in the slowly rotating regime. For the slowly rotating case, perturbative solutions according to the coupling constant, $A$, have been found in the literature \cite{Yagi:2012ya}, while a solution that accounts for all the powers of $\gamma$ can be found in \cite{Chatzifotis:2022mob}. The backreaction to the metric, up to first order in rotation, appears only on the $t\phi$ component and it reads
\begin{equation}
g_{t\phi}(r,\th)=-\left[\frac{r_h}{r}+\sum_{n=4}\frac{d_n}{2^{n-2}}\left(\frac{r_h}{r}\right)^{n-2} \right]\a \sin^2\th~,
\end{equation}
where $\a$ is the rotation parameter (angular momentum per unit mass), while the coefficients $d_n$ are fully determined by the following recurrence relation
\begin{equation}
d_n=\frac{2(n-5)^2(n-1)}{n(n-6)(n-3)}d_{n-1}+2^4\frac{576}{n(n-3)}\g^2d_{n-6}\;,\;\;\;\text{for}\;n\geq 10
\end{equation}
with initial values
\begin{equation}
d_4=d_5=0\;\;,\;\;d_6=-80\g^2\;\;,d_7=-\frac{960}{7}\g^2\;\;,\;\;d_8=-216\g^2\;\;,\;\;d_9=0
\end{equation}
As a final example, we present the so-called extended GB gravity \cite{Fernandes:2021dsb}, which is described by the following action
\begin{equation}
S=\int d^4x\frac{\sqrt{-g}}{2\k^2} \left[ R -\b e^{2\phi}(R+6(\nabla\phi)^2)-2\lambda e^{4\phi}-\tilde{a}\left(\phi\mathcal{R}_{GB}-4G_{\m\n}\nabla^\m\phi\nabla^\n\phi-4\square\phi(\nabla\phi)^2-2(\nabla\phi)^4 \right) \right]~.
\end{equation}
The advantage of this theory lies on the fact that a black hole solution can be obtained analytically. Specifically, for a spherically symmetric and static spacetime, we get
\begin{equation}
-g_{tt}(r)=1+\frac{r^2}{2\tilde{a}}\left[1-\sqrt{1+4\tilde{a}\left(\frac{2M G}{ r^3}+\frac{C}{r^4}\right)}   \right]~,
\end{equation}
for which the event horizon lies on
\begin{equation}
r_h=MG+\sqrt{M^2G^2-\tilde{a}+C}~.
\end{equation}
With the action written in the above form, the coupling constant has dimensions of length square. In order to define the coupling constant to have dimensions of length, we have just to redefine $\tilde{a}\rightarrow A\k$; then, $A$ is the coupling constant of the theory with dimensions of length, as previously. Choosing the following profile for the scalar field \cite{Fernandes:2021dsb}
\begin{equation}
\phi= \ln{\left(\frac{c_1}{r+c_2}\right)}~,
\end{equation}
we get the following constants and parameters of the theory: $C=2\tilde{a},\;c_1=\sqrt{-2\tilde{a}/\beta},\;c_2=0$ and $\la=\b^2/4\tilde{a}$, where $\tilde{a}\b<0$. The above theory has again a smooth limit for a vanishing coupling constant. This allows us to expand the solution in a series expansion arround $\tilde{a}=0$. Then, the solution takes the following form
\begin{equation}
-g_{tt}(r)=1-\frac{r_h}{r}+\left[\frac{r_h}{r}-2\left(\frac{r_h}{r}\right)^2+\left(\frac{r_h}{r}\right)^4\right]\g
-2\frac{r_h}{r}\left[\left(\frac{r_h}{r}\right)^3-2\left(\frac{r_h}{r}\right)^4+\left(\frac{r_h}{r}\right)^6\right]\g^2+\mathcal{O}(\g^3)~.
\end{equation}
Aside from the differences of the above theories and the corresponding hairy black solutions, we are interested in their common features, for which we may state the following
\begin{enumerate}
\item The theories support hairy black hole solutions acquiring a secondary scalar charge, i.e. every dimensionful quantity that parametrises  the solution depends on the ADM mass, the gravitational constant and the coupling constant of the interaction.
\item The corresponding local solutions have a smooth limit for a vanishing coupling constant, which denotes the absence of the interaction.
\item The dependence of the backreacting terms vanish asymptotically, which means that they depend only to inverse powers of $r$
\item  The coupling constant of the interaction appears only into the dimensionless factor $\gamma$ and the $r-$dependence of the backreaction terms appear only into $x=r_h/r$.
\end{enumerate}

In the case where the zero coupling limit $A\rightarrow 0$ corresponds to the Einstein-Hilbert action\footnote{For the extended GB theory~\cite{Fernandes:2021dsb} the absence of the interaction corresponds to the  Einstein gravity with a conformally coupled scalar field.}, the metric has to coincide with the Schwarzchild solution, i.e. $\gamma\rightarrow 0$ (assuming spherical symmetry and staticity). Then, in such a case, the $g_{tt}$ component of the metric would have the following form as a series expansion in $\gamma$
\begin{equation}
	g_{tt}(r)=-(1-x)\left(1+\sum_n\g^{n}f_n(x)\right)~,
	\label{GeneralMetric}
\end{equation}
where $x=r_h/r$ and $f_n(x)$ some polynomial functions. The exterior of the black hole, $r>r_h$ corresponds to the interval $0<x<1$, where $x\rightarrow 0$ corresponds to $r\rightarrow\infty$ and $x=1$ to the horizon $r=r_h$. Thus, asymptotic flatness implies that $f_n(0)=0$, i.e. there is no constant term to the polynomials.

Moreover, at the limit of large black holes, the backreacting terms relax to zero (as $\g\rightarrow 0$ via its denominator) and the metric is effectively reduced to the Schwarzschild black hole. At the exterior, the temporal component has to be negative, except for the horizon that has to be zero, which is guaranteed by the Schwarzschild term, $(1-x)$.
This implies,
\begin{equation}
	1+\sum_{n}\g^{n}f_n(x)>0,\;\;\text{for}\;\;0< x< 1~.
\end{equation}
In the interior of the black hole, $x>1$, the aforementioned term may in principle vanish, leading to an inner horizon. Thus, the condition for the existence of an inner horizon reads
\begin{equation}
	1+\sum_n\g^{n}f_n(x_0)=0,\;\;\text{for some}\;\; x_0> 1~.
	\label{innerhorizon}
\end{equation}
What we want to investigate, is whether the interactions between the gravitational field and the matter fields content may lead to a stable black hole branch. Assuming the case where (\Ref{innerhorizon}) has at most one physically acceptable root,  there might be two asymptotic limits for the black hole in the $\tau-r_h$ plane. These asymptotic limits correspond to the horizon radius for which $\tau\rightarrow\infty$, i.e. to a vanishing temperature. In agreement with the above considerations, we expect that one asymptotic limit corresponds to the case of the large black hole size, which is effectively the Schwarzschild black hole (plus a perturbation due to the interaction), which is the unstable branch. On the other hand, the stable black hole branch exists if there exists a finite (non-zero) $r_h$ for which $\tau\rightarrow\infty$, since, if this is the case, a branch for which $\partial r_h/\partial T>0$ in the $r_h-\tau$ plane has to be there, which implies a positive heat capacity and a corresponding positive topological charge, as was shown in (\Ref{topcharge}). Thus, the existence of the stable branch can be studied in terms of the vanishing limits for the surface gravity. The surface gravity for a spherically symmetric metric is given by
\begin{equation}
	\k_g=\frac{ \vert g_{tt}^\prime\vert   }{2\sqrt{-g_{tt}\, g_{rr}} }\Big\vert_{r\rightarrow r_h}~,
\end{equation}
where prime denotes differentiation in respect to the variable $r$. One has to guarantee that the metric determinant has to be finite for $0<x<-\infty$. This is achieved by imposing that the term $g_{tt}\, g_{rr}$ is finite.  Thus, the surface gravity tends to zero, for those $r_h$ for which $\vert g_{tt}^\prime(r_h)\vert\rightarrow 0$.
For the general metric component \eqref{GeneralMetric}, we get
\begin{equation}
	g_{tt}^\prime(r_h)=-\frac{1}{r_h}\left( 1+\sum_n\g^{n}f_n(1)  \right)~.
\end{equation}
Thus, $g_{tt}'(r_h)\rightarrow0$ for $r_h\rightarrow\infty$, which accounts for the Schwarzschild limit (free gravitational field) and for those finite $r_h$ that satisfy:
\begin{equation}
	1+\sum_{n}\g^{n}f_n(1)=0~.
	\label{StableBranchCondition}
\end{equation}
This condition describes the limit where the event horizon and the inner horizon match at $r=x_0r_h$, i.e. to the limit where the condition \eqref{innerhorizon} is satisfied for $x_0\rightarrow1$. Thus, if the interaction between matter and gravity produces an inner horizon for the black hole, a stable branch appears. The implication of this is that for the case of small black holes, where the perturbation theory on the $\gamma$ parameter breaks down, the hair contribution on the metric ceases to be subdominant and may allow for a first order phase transition on the thermodynamic configuration space under (\Ref{StableBranchCondition}), which in terms implies a jump in the thermodynamical topological sectors from the negative to the positive topological charge of (\Ref{topcharge}). \\

As a final note, we would like to consider the inverse procedure to verify our main result. We will proceed to show that if (\Ref{StableBranchCondition}) applies, then the topological charge will consequently always be positive. Let us consider a hairy black hole metric, whose $g_{tt}$ component is given via (\Ref{GeneralMetric}). In particular, one may consider the following general metric ansatz for a hairy black hole:
\begin{equation}
	ds^2=-F(x)dt^2+\frac{r_h^2}{x^4}\frac{dx^2}{H(x)}+\frac{rh^2}{x^2}d\Omega^2
\end{equation}
extracted by a simple coordinate transformation $\displaystyle x=\frac{rh}{r}$ from the Schwarzschild coordinates. Additionally, following (4.16), one may set that
\begin{equation}
	F(x)=(1-x)\left(1+\sum_{n}\gamma^n f_n(x)\right),\qquad H(x)=(1-x)\left(1+\sum_{n}\gamma^n h_n(x)\right)
\end{equation}
where, again, asymptotic flatness requires that 
\begin{equation}
	\label{flat}
	\sum_{n}\gamma^n f_n(0)=\sum_{n}\gamma^n h_n(0)=0.
\end{equation}
Enforcing (\Ref{StableBranchCondition}), the surface gravity vanishes, which implies that $\tau\rightarrow \infty$ for a finite (non-zero) $r_h$, that is the minimum horizon radius value, such that the free energy becomes on-shell. Then, the $\displaystyle \frac{S}{\tau}$ term of (\Ref{FreeEnergy}) is subdominant, assuming the natural fact that the entropy will be always finite for a finite horizon radius. Arbitrarily close to such an extreme case, the free energy of the system is effectively reduced to the ADM mass of the black hole, which, due to asymptotic flatness, is the same as the Komar mass
\begin{equation}
	M =-\frac{1}{8\pi G} \lim_{x\rightarrow 0} \oint_{S^2} \nabla^\mu K^\nu dS_{\mu\nu},
\end{equation}
where $K^\nu$ is the Killing vector associated with time-translation symmetries, while $dS_{\mu\nu}$ is the reduced volume form on the 2-sphere at infinity. Then, plugging our metric ansatz, it is easy to verify that
\begin{equation*}
	M=\lim_{x\rightarrow 0}  \frac{r_h}{2G}\frac{g'_{tt}(x)}{\sqrt{-g_{tt}(x)g_{rr}(x)}}=\lim_{x\rightarrow 0}\frac{r_h \left((x-1) \sum _{n=1}^{\infty } \gamma ^n f'_n(x)+\sum _{n=1}^{\infty } \gamma ^n f_n(x)+1\right)}{2 G \sqrt{\frac{\sum _{n=1}^{\infty } \gamma ^n f_n (x)+1}{\sum _{n=1}^{\infty } \gamma ^n h_n(x)+1}}}
\end{equation*}
where prime denotes differentiation with respect to $x$. Imposing (\Ref{flat}), we find that
\begin{equation}
	\label{adm}
	M(r_h)=\frac{r_h}{2G}(1-\sum _{n=1}^{\infty } \gamma ^n f'_n(0)),
\end{equation}
We note that, for a physically acceptable solution, the ADM mass has to be positive, which imposes the inequality
\begin{equation}
	\sum_{n=1}^{\infty}f'_n(0)\gamma^n<1,\qquad \forall r_h.
\end{equation}
Under the paradigm of $\mathcal{F}\approxeq M$, the condition for a thermally stable configuration simply reads that
	\begin{equation}
	\frac{\partial\mathcal{F}}{\partial r_h}\Big|_{r_h=r_h^*}\approxeq\frac{\partial M}{\partial r_h}\Big|_{r_h=r_h^*}=0,\qquad \frac{\partial^2 \mathcal{F}}{\partial r_h^2}\Big|_{rh=rh^*}\approxeq\frac{\partial^2 M}{\partial r_h^2}\Big|_{rh=rh^*}>0
	\end{equation}
where $r_h^*$ corresponds to the on-shell black hole horizon radius. Essentially, the problem reduces to the following statement: Given the positive ADM mass of (\Ref{adm}), assuming that it contains at least one local extremum (or in principle more), we need to show that the smaller on-shell radius, i.e. for which the ADM mass yields an extremum, corresponds to a local minimum. This is actually easily verified from the very form of the ADM mass function. Indeed, we note that (\Ref{adm}) can be expressed as
\begin{equation}
	M(r_h)=\frac{r_h}{2G}-\frac{r_h}{2G}\mathcal{O}\left(\frac{A\kappa}{r_h^2}\right)-\frac{r_h}{2G}\mathcal{O}\left(\frac{A\kappa}{r_h^2}\right)^2-...,
\end{equation}
where we used $\displaystyle \gamma=\frac{A\kappa}{r_h^2}$. Therefore, 
\begin{equation}
	M(r_h)\rightarrow \infty,\qquad (\forall r_h\rightarrow 0)\,\text{and}\,( \forall r_h\rightarrow\infty)   
\end{equation}
This implies that, from basic arguments on continuity and differentiability that, if a local extremum of $M(r_h)$ exists, then the lowest local extremum $r_h^*$ we obtain will be a local minimum with $\displaystyle \frac{\partial^2 M}{\partial r_h^2}\Big|_{rh=rh^*}>0$. This configuration, under the condition of vanishing temperature, implies that
\begin{equation}
	\frac{\partial^2 \mathcal{F}}{\partial r_h^2}\Big|_{rh=rh^*}>0,
\end{equation}
which, in view of equation (\Ref{topcharge}), corresponds to a positive topological charge. This solidifies our understanding that if the black hole is allowed to become extremal due to the contribution of secondary hair, the final thermal stage of the black hole will always be reached through a stable branch of positive winding number.   
The careful reader shall notice that this procedure breaks down for the case of $\sum_{n=1}^\infty f'_n(0)\gamma^n=0$. Indeed, for this case, the ADM mass is the same as the Schwarzschild mass and thermal stabilization due to the hair contribution cannot be achieved, which is a result that was missed in the prior analysis. Naturally, in order to avoid this, $f_n(x)$ need to be polynomials that contain at least one term linear in $x$. 

\subsection{ Black Hole stability and a potential minimum length in quantum spacetime }

The condition \eqref{StableBranchCondition} implies a lower limit for the black hole size that can be in thermal equilibrium with the heat bath. This lower limit is given in terms of the coupling constant of the interaction, through \eqref{StableBranchCondition}, and is the lower size limit for which the black hole tends to acquire a zero temperature; in other words, the case for which the black hole tends to be extremal. The condition \eqref{StableBranchCondition} is acquired for some critical value of the $\gamma$-factor. We can measure the coupling constant $A$ in terms of the Planck length, $l_P =\kappa/\sqrt{8\pi}$, which means that we can introduce a dimensionless parameter $\widetilde{A}$, in such a way that $A=\widetilde{A}^2l_P$. Then, the lowest possible size of the black hole horizon is of the following order of magnitude
\begin{equation}\label{rhA}
r_{h,min}\sim |\widetilde{A}| \,l_P~.
\end{equation}
From the explicit examples of black holes we have examined above,
it becomes clear that $A$ depends on the parameters and couplings of the underlying microscopic theory of gravity and therefore, whether the stable black hole minimal size exceeds or not the Planck length depends on the order of magnitude of the coupling of the pertinent interaction.

Hawking radiation is one of the main ways a black hole evolves down to lower sizes. If the interaction produces an inner horizon in the black hole, then the flow of thermal energy away from the black hole comes to an end. Therefore, if an inner horizon exists, as is the case of our specific examples of Kerr and RN black holes examined in section \ref{sec:bath} ({\it cf.} fig.~\ref{bhbranches}), then,  at the point of maximum temperature $T_{\rm max}$ (occurring at $\partial r_h/\partial \tau = \infty, \, \tau =T^{-1}$),  one observes that the heat capacity \eqref{cq2} diverges at $T=T_{\rm max}$, and subsequently changes sign, thus signalling a first-order phase transition. Such a first-order transition is also in correspondence with our earlier topological considerations associated with the
discontinuous change of the winding number from  $-1$ to $+1$.
Given \eqref{rhA}, the minimum horizon size induced by the phase transition should be larger than the Planck scale for
\begin{align}\label{transpl}
A/\ell_P  \ge 1\,,
\end{align}
if subplanckian horizon sizes are to be avoided.\footnote{For completion, we mention that, in our specific examples within the string theory framework, subPlanckian horizon sizes are avoided, for the case of Gauss Bonnet interactions alone, (\eqref{strings}), for weakly coupled strings $g_s \ll 1$ even if $\alpha^\prime \sim \kappa^2$, while for the corresponding Chern-Simons couplings \eqref{stringsCS}
subplanckian horizon sizes are excluded for $\alpha^\prime \gtrsim \kappa^2$.} As a result, a thermally stable black hole remnant can occur.

The above concept of the production of an inner horizon determined by a specific length scale introduced into the theory, is not only a feature of appropriate interactions between gravity and matter fields with dimensionful couplings. It is also introduced in theories with a fundamental physical length scale. In \cite{Nicolini:2008aj} a length scale $\theta$ introduced through the Generalized Uncertainty Principle (GUP), in such a way that it regularises the singularity of a black hole, while also leads to a similar thermodynamic behaviour as the one described above,  to the production of an inner horizon,  signalling the end of the runaway evaporation of the black hole. As such, the introduction of quantum effects, either in the form of an effective theory and interactions between gravity and matter fields, or through principles of Quantum Mechanics that we assume to hold for the spacetime, like the GUP, are able to stabilise thermodynamically the black hole at the late stages of its evaporation through the production of an inner horizon inside the event horizon.

It is understood that our semiclassical considerations are not valid when the size of the minimum inner horizon approaches the quantum gravity limit of the Planck length, nevertheless the whole approach makes it plausible to conjecture a potential connection of the thermodynamical criterion of \cite{Wei:2022dzw} with the field of quantum gravity, in the sense that the existence of thermally stable microscopic black holes with minimal horizons of Planck size might constitute the structure of space time foam itself~\cite{Wheeler:1955zz}.

In the above examples of the linear-dilaton-GB, or Chern-Simons gravity, we have dealt with analytic expressions for the corresponding black-hole solutions, in which  
the backreaction of the hair field to the spacetime geometry is known only up to first orders w.r.t. the coupling constant.
Unfortunately this is not the case for the full stringy case of the black hole with secondary dilaton hair of  \cite{Kanti:1995vq}, characterised by exponential dilaton couplings to the GB curvature combinations.  Thus, the behaviour of the system for smaller sizes of the black hole cannot be deduced in an analytic fashion. In the case of the GB interaction some conclusions have been argued, 
in the recent literature, but under some special circumstances that seem to be vague. Firstly, in \cite{Kanti:1995vq}, perturbative solutions near the black-hole horizon have been given in the form of polynomial expansions in integer positive powers of $(r-r_h)$. In \cite{Kanti:1996gs} the near horizon and asymptotic solution of the theory reveal a lower size limit for which a black hole solution can exist, but {\it only} under the assumption that the aforementioned coefficients were independent of  $r_h$. Unfortunately, the coefficients of such expansions are themselves functions of the horizon-radius size $r_h$, which prevents one from making generic analytic statement on the existence of a lowest size for which such black hole solutions can exist. In order to verify that such constants are indeed functions of $r_h$, consider the generic metric ansatz \eqref{GeneralMetric} and Taylor expand it around $x=r_h/r=1$, i.e. near the horizon. Then, the $g_{tt}$ component at the near horizon regime has the following form:
\begin{equation}
	g_{tt}(r\approx r_h)=-\frac{1+\sum_n \gamma^n f_n(1)}{r_h}(r-r_h)+\mathcal{O}((r-r_h)^2)
\end{equation}
As such, the coefficient in front of $(r-r_h)$ is indeed a function of $r_h$ due to both, the denominator, but also, and more inportantly, to the numerator, which owes its existence to the backreaction. It is important to note that the numerator vanishes if the condition \eqref{StableBranchCondition} for the appearance of the stable branch  is true for some critical value of $\gamma$. This means that, if an inner horizon exists, the near horizon approximation of \cite{Kanti:1996gs} totally breaks down when the event horizon tends to coincide with the inner horizon, which is reasonable because a near horizon approximation does not take into account the black hole's interior. In addition, it is well known that extremal black holes have a topology of $AdS^2\times S^2$ near the horizon regime, which in turn raises questions about the validity of the near horizon solution. Moreover, in \cite{Corelli:2022phw} a numerical approach is considered in order to investigate the highly non-linear effects of the GB coupling in the string-inspired case of \cite{Kanti:1995vq}, albeit also assuming constant coefficients in the expansion around the horizon $r_h$. Such simulations reveal that indeed a stable branch appears at the final stages of evaporation, without, however, an extremal black hole of zero temperature as an endpoint, which could be considered unphysical, since, if true, it would imply either a violation of the first law of thermodynamics or a fixed black-hole entropy, which is equally strange. We expect that such results are misleading for the fate of the dilatonic black hole, due to the failure of the numerical method, since the coefficients are horizon dependent in a non-trivial way. However, according to the considerations of the present work, the existence of a minimum black hole size is closely related to the appearance of a stable branch for the black hole. In this sense, the numerical methods might indeed be true about the existence of a minimum size black hole, but the non-zero temperature at this critical point might be a misleading result due to the failure of the near horizon approximation scheme, which as we already stated totally fails close to the minimum size.  
Of course, lacking an exact analytic solution in the model of   \cite{Kanti:1995vq} prevents one from reaching definitive conclusions on this issue.  To extend the analysis of \cite{Corelli:2022phw} to include such a case would require the imposition of appropriate boundary conditions at the origin along with precision method of numerical relativity, as well as an appropriate choice for the range of the pertinent Gauss Bonnet coupling. Such a non-trivial analysis falls way beyond the scope of this paper.
\\
\\
\\
\\

\section{Conclusions and Outlook}\label{sec:concl}

In this work, we have elaborated further on the conjecture of \cite{Wei:2022dzw}, which concerns
a criterion for the thermodynamical stability of a black hole viewed as a defect in thermodynamic space by means of the positivity of an appropriately defined winding number. We have managed to verify  the conjecture in some cases of black hole with secondary scalar hair that backreacts on the geometry.  Our analysis involved examples of (pseudo)scalar hair induced by interactions of the matter fields with quadratic-curvature combinations, which are characterised by dimensionful couplings with mass dimension $-1$. We have argued on the applicability of the criterion of \cite{Wei:2022dzw} to these examples, which admit analytic perturbative solutions in the coupling. These include gravitational theories with linear dilaton-Gauss-Bonnet coupling, Chern-Simons gravity with pseudoscalar fields coupled linearly to the gravitational anomaly, and also the extended Gauss-Bonnet gravity, which also involves scalar fields linearly coupled to the Gauss-Bonnet invariant. All such theories admit black holes with (pseudo)scalar secondary hair.

We have paid particular attention to discussing the r\^ole of the back reaction of the hair matter fields onto the spacetime geometry on the stability of the black holes. The strength of the back reaction is determined
by appropriately dimensionless combinations  involving the interaction of coupling of matter with the geometry,  which has dimensions of length (we consider examples in which such an interaction corresponds to a
linear coupling of the scalar field to curvature squared combinations). An important aspect of the back reaction is the potential appearance of an inner horizon, which in the case of thermal stability coincides with the event horizon, thus leading to extremality. It is important to notice that because of the     secondary nature of the hair, which by the way respects the spirit of the no-hair conjecture, such contributions on the background geometry cannot be shed away during Hawking evaporation.

We have demonstrated that such dimensionful couplings imply the potential existence of minimal inner horizons for the black holes that are thermodynamically stable, in the sense of the aforementioned stability criterion. To avoid black holes with sizes below the Planck length (a sort of transplanckian conjecture) one should obviously impose certain conditions on the size of the interaction of the higher curvature coupling, which should be larger than the Planck length. This in turn imposes restrictions on the range of the fundamental parameters of the microscopic theory which enter the expression for this coupling. The existence of such minimal-size stable black hole remnants lead us to conjecture that the quantum spacetime structure of the theory consists of space time foam,  involving (quantum fluctuating) microscopic black holes of Planck size.

An interesting possible direction would be to apply the above ideas so as to understand stable dark matter remnants provided, for instance, by primordial black holes in modified gravity theories involving scalar fields coupled to higher curvature interactions. Moreover, spacetimes characterised by non constant surface gravity, such as Vaidya spacetimes, also constitute interesting arenas where the topological stability conjecture can be tested.

\section*{Acknowledgments}

The work of N.C. and E.P is supported by the research project of the National Technical University of Athens (NTUA) 65232600-ACT-MTG:  {\it Alleviating Cosmological Tensions Through Modified Theories of Gravity}; that of P.D. is partially supported by
a NTUA scholarship, while the work of N.E.M. is supported in part by the UK Science and Technology Facilities research Council (STFC) and UK Engineering and Physical Sciences Research
Council (EPSRC) under the research grants ST/T000759/1 and  EP/V002821/1, respectively. N.E.M.  also acknowledges participation in the COST Association Action CA18108 ``{\it Quantum Gravity Phenomenology in the Multimessenger Approach (QG-MM)}''.

\end{document}